\documentclass[aps,prl,letterpaper,11pt,twoside,tightenlines,nofootinbib,showpacs,preprint,twocolumn]{revtex4}
\usepackage{graphicx}
\usepackage[sort&compress]{natbib}
\usepackage{latexsym}
\usepackage{epsfig}
\usepackage{amsmath}

\def\q{q \bar q}

\def\be{\begin{equation}}
\def\ee{\end{equation}}

\def\NP{{ Nucl.\ Phys.\ }}
\def\PL{{ Phys.\ Lett.\ }}
\def\PR{{ Phys.\ Rev.\ }}

\begin{document}

\title{Confinement Horizon and QCD Entropy}
\author{Paolo Castorina$^{1,2}$ and
Alfredo Iorio$^{\rm 3}$}
\affiliation{
\mbox{${}^1$ Dipartimento di Fisica, Universit\`a di Catania, Via Santa Sofia 64,
I-95123 Catania, Italy}\\
\mbox{${}^2$ INFN, Sezione di Catania, I-95123 Catania, Italy.}\\
\mbox{${}^3$ Institute of Particle and Nuclear Physics, Faculty of Mathematics and Physics, Charles University}\\
\mbox{V Hole\v{s}ovi\v{c}k\'ach 2, 18000 Prague 8, Czech Republic}
}

\date{\today}
\begin{abstract}
Within the picture of quark confinement as due to a \textit{color event horizon}, and of hadronization as an instance of the Unruh radiation for the strong force, we show here that QCD entropy, evaluated by lattice simulations in the region $T_c < T < 1.3 T_c$, is in reasonable agreement with a melting color event horizon.
\end{abstract}
 \maketitle

\section{Introduction}

The interpretation of quark confinement as the effect of a classical \textit{event horizon} for color degrees of freedom \cite{Casto,Salam}, naturally lead to view hadronization as the quantum tunnelling through such horizon \cite{K1,K2}. From this point of view, hadron formation is the result of the \textit{Unruh radiation} \cite{Un} associated to the strong force.

More precisely, hadronization is the result of a Unruh phenomenon related with the string breaking/formation mechanism, that is, with the large distances QCD behavior.
The hadronic-Unruh temperature is given by \cite{K2}
\be
T_{h} = \frac{a}{2 \pi} \simeq \sqrt{\frac{\sigma}{2 \pi}}
\label{0}
\ee
where $\sigma$ is the string tension, and the acceleration, $a \simeq 2 k_T \simeq \sqrt{2\pi \sigma}$, is the one necessary to bring on shell a quark of transverse momentum $k_T$. In other words, the time given by the characteristic fluctuations determined by the virtuality of the pair is $\Delta \tau = 1/\Delta E \simeq 1/(2 k_T)$.

Moreover, at zero chemical potential, $\mu = 0$, one has $T_{h} = T_c$, the critical temperature of the deconfinement transition, because it is directly related with the string breaking/formation mechanism. This provides a theoretical basis for understanding the production of newly formed hadrons in high energy collisions, and, as shown in \cite{CIS1}, it allows to predict ($\mu=0$): a) the hadronic freeze-out conditions \cite{taw,jean1,jean2},
i.e. $s/T_{h}^3 = 3 \pi^2 /4 \simeq 7.4$ in terms of the entropy density $s$, and b) the value $\langle E \rangle / \langle N \rangle = \sqrt{2 \pi \sigma} \simeq 1.09$
for the average energy per hadron.  These predictions  are based on the string breaking/formation mechanism, and on the adaptation of the Bekenstein-Hawking (BH) entropy formula \cite{bekensteinEntropy} to the color event horizon \cite{CIS1}. The BH formula was born in the context of black hole physics, however, there is by now a vast literature where its implications are studied in more general contexts, see, e.g.,\cite{bousso}. Some connections between the hadronization as a Unruh phenomenon and the corresponding near horizon black hole scenarios, have been studied in \cite{CGI}.

Although the Unruh hadronization mechanism holds at the string breaking, that is, at the corresponding $T_c$, since the deconfinement transition is a cross-over, one can expect some remnant of confinement slightly above $T_c$. Indeed the persistence of string-like objects above $T_c$ has ben obtained by many different methods: lattice simulations \cite{karsch,cea}, quasiparticle approach \cite{mannarelli1,mannarelli2}, NJL correlator \cite{beppe,jap}, Mott transitions \cite{david} and confinement mechanisms \cite{eddy}. It is then natural to ask whether small changes in the description of color confinement, slightly above $T_c$, can give information on thermodynamical quantities, as, e.g., the QCD entropy.

In this paper we show that the QCD entropy, evaluated by lattice simulations \cite{olaf}, in the region  $T_c < T < 1.3 T_c$, is in reasonable agreement with the picture of a melting color event horizon.

Next we recall more details of the Unruh hadronization mechanism, including the understanding, in terms of string-breaking, of the other freeze-out condition: $n \simeq 0.12$ fm$^{-3}$, where $n$ is the number density. We then discuss temperature effects related to the string-like system near the phase transition, give our results for the entropy and the internal energy as compared with lattice QCD simulations, and draw some conclusions.

\section{The Unruh hadronization}

%\subsection{Unruh hadronization temperature}

\begin{figure}[h]
\centerline{\psfig{file=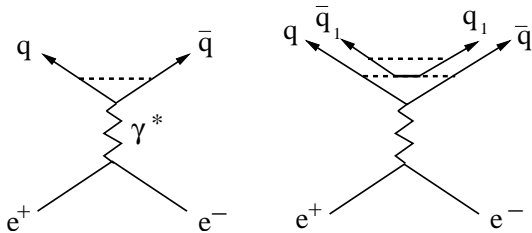,width=7cm} }
\caption{Quark formation in $e^+e^-$ annihilation}
\label{anni}
\end{figure}

Although universal, the mechanism is most simply illustrated by hadron production through
$e^+ e^-$ annihilation into a $\q$ pair, as shown in Fig.\ref{anni}.

The attempt to separate the initial $\q$ pair  ends at a distance $R$, when both the quark and the antiquark hit the confinement horizon, that is, when they reach the end of the binding string. The separation can now continue only if a further quark-antiquark system, say $q_1 \bar{q}_1$, is excited from the vacuum. Although the new pair $q_1 \bar{q}_1$ is at rest in the overall center of mass, each of its constituents has a transverse momentum $k_T$, determined by the uncertainty relations in terms of the transverse dimension of the string flux tube. String theory gives for the basic thickness \cite{Lue}
\be
r_T=\sqrt{2/\pi \sigma},
\label{1}
\ee
leading to
\be
k_T=\sqrt{\pi \sigma/2}.
\label{2}
\ee
The maximum separation distance $R$ can thus be obtained from $\sigma R = 2  k_T$, hence, from (\ref{1}) and (\ref{2}) one has
\be
R = \sqrt{2 \pi / \sigma} .
\label{4}
\ee

% \subsection{Entropy of color event horizon}

The entropy associated to a color event horizon is necessarily an entropy of entanglement, between quantum field modes on both sides of the horizon. Its general form is \cite{terashima,entaentropy}
\be
\label{entropymatter}
S_{\rm ent} = \alpha \frac{A_h}{r^2} \;,
\ee
where $A_h$ is the area of the event horizon, $r$ the scale of the characteristic quantum fluctuations, and $\alpha$ an undetermined numerical constant. This expression shares its holographic
behavior\footnote{Holography of entanglement entropy is a quite general
result, see \cite{srenidcky,solodukin}.} with the BH entropy formula of a black hole \cite{bekensteinEntropy}
\be
S_{\rm BH} = {1 \over 4}~ {A \over r_P^2} ,
\label{10}
\ee
where $A$ denotes the surface area of the hole, that, e.g., in the Schwarzschild case is given by $A = 4 \pi R_S^2$, with $R_S=2GM/c^2$. The quantity $r_P= \sqrt{\hbar G/c^3}$
is the Planck length the smallest possible fluctuation scale.

As shown in \cite{laflamme}, a formula similar to the Bekenstein-Hawking formula (\ref{10}) holds in the case of the Rindler spacetime of an accelerated observer. On the other hand, it is well known that the Rindler spacetime can be associated to the near-horizon approximation of a black hole spacetime \cite{wald}

In ref. \cite{CIS1}, the above has been applied to the Unruh hadronization mechanism, allowing to predict within the model the freeze-out conditions.
Indeed, in this case the characteristic scale of the quantum fluctuations is given by Eq.\ (\ref{1}), and we obtain
\be
S_h = {1\over 4}~ {A_h \over r_T^2} = {1\over 4}~ {4 \pi R^2 \over r_T^2} = \frac{\pi^2}{2} \sigma R^2
\label{11}
\ee
for the entropy associated to hadron production. The parameter $R$ is given by Eq.\ (\ref{4}), and the smallest
fluctuation scale is the transverse string thickness (\ref{1}). Using Eq.\ (\ref{4}) into Eq.\ (\ref{11}) gives
\be
S_h = \pi^3,
\label{S}
\ee
while the entropy {\sl density} divided by $T^3$, evaluated at ${T=T_c}$, gives
\be
{s  \over  T^3} = {S_h \over (4 \pi/3) R^3 T^3} =
{3 \pi^2 \over 4} \simeq 7.4
\label{entrop}
\ee
as freeze-out condition in terms of $s(T)$ and $T$. This result is in agreement with the value obtained for $s/T^3$ from species abundance analyses
in terms of the ideal resonance gas model \cite{Cley,Tawfik}.

Furthermore, one can shows that the other freeze-out condition, based on the number density $n$, that is $n \simeq 0.12$ fm$^{-3}$ \cite{foc} is directly related with the string-breaking too.
Indeed, for a single string-breaking the number density is given by
\be
n_{sb}= \frac{1}{4\pi R^3/3}
\ee
where $R$ is the string breaking distance, which turns out to be $R=1/T_h$, for massless quarks. For $T_h \simeq 160$ MeV, one obtains $n_{sb} \simeq 0.129$ fm$^{-3}$.

From the above, it should be clear that the previous formulae hold strictly at the string breaking, that is, at the hadronization temperature $T_c$. Therefore, to compare results obtained in this approach with lattice data at $T\ne T_c$ requires a more general analysis that we now present.

\section{Color horizon entropy slightly above $T_c$}

A natural starting point is to generalize Eq.\ (\ref{11}) and write
\be
S_h (T) =  \frac{\pi^2}{2} \sigma(T) R^2(T) ,
\label{ShT}
\ee
where $\sigma(T)$ is the string tension for $T \ge T_c$, which for a sharp deconfinement transition should be exactly zero, and $R(T)$ has to be interpreted as the effective range of the color field above $T_c$.

\begin{figure}
{{\epsfig{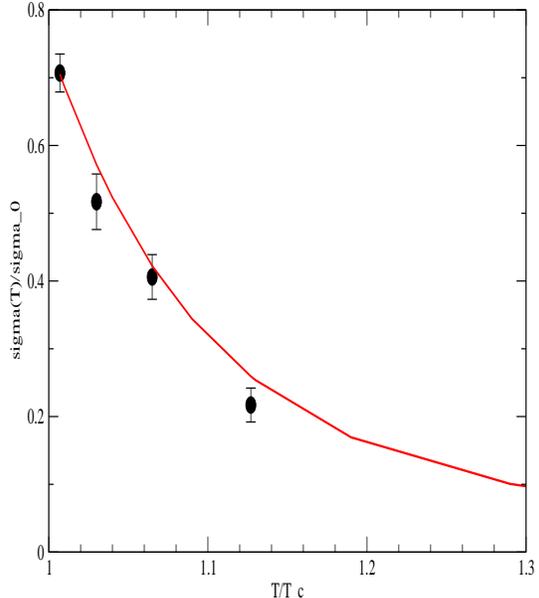}}
\caption{Behavior of the string tension $\sigma(T)$ above $T_c$.
}
\label{Fig:sigmabeyondTc}}
\end{figure}

For the observed crossover between the quark-gluon phase and the hadron phase one should expect that $\sigma(T)$ and $R(T)$ go quickly to zero. Some information about the behaviour of $\sigma$ above $T_c$ can be obtained as follows. The string tension $\sigma$ can be interpreted as the vacuum energy density in a flux tube of transverse area $\pi r_T^2$, i.e.
\be
\sigma = \epsilon_v \pi r_T^2 .
\label{vacuumsigma}
\ee
On the other hand, from Eq.\ (\ref{1}) we have $r_T^2 \simeq 1/\sigma$, hence the string tension scales with the square root of the vacuum energy $\sigma \simeq \epsilon_v^{1/2}$.

The behaviour of the vacuum energy density of the chromoelectric field above $T_c$ has been evaluated in lattice QCD in ref. \cite{adriano1,adriano2}, where  the ratio of the vacuum energy density at high temperature $T>T_c$, to its value for $T<T_c$ is given. From the previous discussion, the behavior of the string tension above $T_c$ is given by
\be
\frac{\sigma(T)}{\sigma(T_c)} = \left( \frac{\epsilon_v(T)}{\epsilon(T_c)} \right)^{1/2} ,
\label{sigmaTsigmaTc}
\ee
and it is depicted in Fig.\ref{Fig:sigmabeyondTc}, by using the results in Ref.\cite{adriano2} and a fit (red curve), where $\sigma_0$ is the string tension at $T_c$ (below $T_c$ the chromoelectric field is essentially constant).

\begin{figure}
{{\epsfig{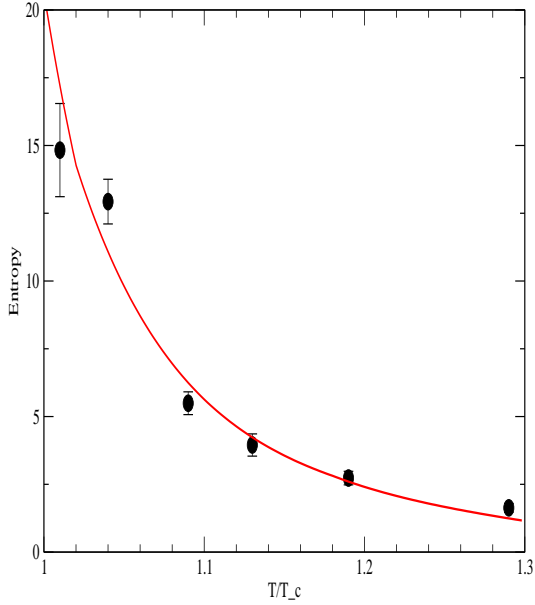}}
\caption{Entropy at large distances, $S_h(T)$, vs temperature. Here $R(T)/R_0 = a + b \exp \left[ -c ( T/T_c -1) \right]$, and $a+b=1$ and $a \simeq 0.3$, and $S_h$ multiplied by $2/3$ to compare with lattice data on 2-flavour QCD.}\label{Fig:Sh2/3}}
\end{figure}

Using Eq.\ (\ref{sigmaTsigmaTc}) in Eq.\ (\ref{ShT}), the entropy at large distances can be evaluated by the equation
\be
S_h(T) =  \pi^3  \frac{\sigma(T)}{\sigma_0} \left( \frac{R(T)}{R_0} \right)^2 ,
\label{ShT2}
\ee
where $R_0 = R(T_c)$. To compare with lattice data one needs the ratio $R(T)/R_0$ which we parametrize as
\be
R(T)/R_0= a + b \exp \left[ -c ( T/T_c -1) \right] ,
\ee
where $ a+b=1$ and $a \simeq 0.3$ has been fixed by the  $T$-independent gluon field correlation length \cite{adriano1,adriano2}  $\ell =  a R_0\simeq 0.34$fm with $R_0 \simeq 1.1$ fm.
To compare with lattice data on 2-flavour QCD one has to multiply $S_h$ in Eq.\ (\ref{ShT2}) by $2/3$. The results are in Fig.\ref{Fig:Sh2/3}.

\begin{figure}
{{\epsfig{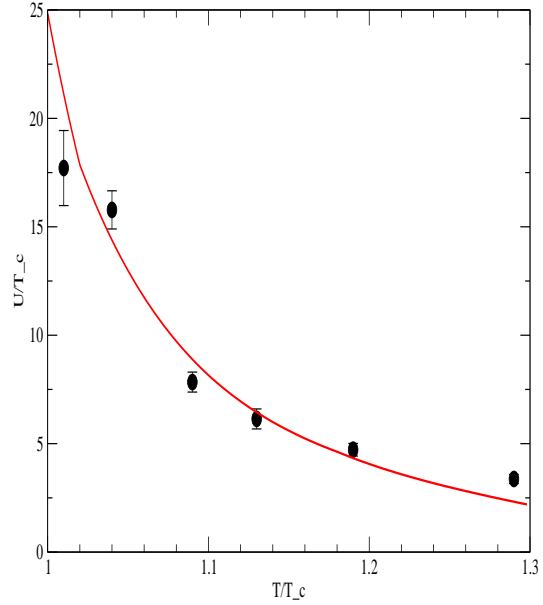}}
\caption{Internal energy at large distances, $U_\infty$, vs temperature.}\label{Fig:intener}}
\end{figure}

In the proposed framework, a calculation of the internal energy at large distances, $U_\infty$, is possible by considering that $U = F + T S$, and that $F= \sigma(T) R(T)$, see Eq.\ (\ref{vacuumsigma}). Thus
\be
U_\infty  = F + T S_\infty  =  \sigma(T) R(T) + T S_\infty .
\ee
From the above it turns out
\begin{eqnarray}
\frac{U_\infty}{T_c} & = & \frac{\sigma(T)}{\sigma_0} \frac{R(T)}{R_0} \nonumber \\
&& \times \left( 2\pi + \pi^3 \frac{T}{T_c} \frac{R(T)}{R(0)} \right) .
\end{eqnarray}
and the comparison with the lattice data requires again the factor 2/3. The results are in Fig.\ref{Fig:intener}.

\section{Conclusions}

The consistency between our results and lattice data on the QCD entropy above the critical temperature suggests the picture of a progressive melting of the
color confinement horizon. This dynamical  decription is completely in agreement with the persistence of string-like structures that survive slightly above $T_c$.

\section*{Aknowledgements}

The authors thank Helmut Satz and Martin Spousta for useful discussions. P.C. gladly acknowledges the kind hospitality of the Institute of Particle and Nuclear Physics, MatFyz, of Charles University, where this work was initiated.


\begin{thebibliography}{99}

\bibitem{Casto} E.\ Recami and P.\ Castorina, Lett. Nuovo Cim. 15 (1976) 347.

\bibitem{Salam} A.\ Salam and J.\ Strathdee, \PR D18 (1978) 4596;\\
see also C.J.\ Isham, A.\ Salam, and J.\ Strathdee, \PR D3 (1971) 867.

\bibitem{K1}
  D.~Kharzeev, E.~Levin and K.~Tuchin,  Phys.\ Rev.\  C {\bf 75}, 044903 (2007).

\bibitem{K2} P.Castorina, D. Kharzeev and H. Satz Eur.Phys.J. C52 (2007) 187-201.

%\bibitem{Haw} S.\ W.\ Hawking, Comm.\ Math.\ Phys.\ 43 (1975) 199.

\bibitem{Un} W.\ G.\ Unruh, \PR D14 (1976) 870.

\bibitem{CIS1} 	P.~Castorina, A.~Iorio and H.~Satz, Int. J. Mod. Phys. E24 (2015) 1550056.

\bibitem{jean1} J. Cleymans and K. Redlich, Phys. Rev. Lett. 81 (1998) 5284; J. Cleymans and K. Redlich, \PR C61 (1999) 054908;J. Cleymans et al., arXiv:hep-ph/0511094.

\bibitem{taw} A. Tawfik, J. Phys. G 31 (2005) S1105; hep-ph/0507252 and hep-ph/050824.

\bibitem{jean2}  V. Magas and H. Satz, Eur. Phys. J. C32 (2003) 115; J. Cleymans et al., Phys. Lett. B 615 (2005) 50.

\bibitem{bekensteinEntropy} J.~D.~Bekenstein, \PR D23 (1973) 2333.

\bibitem{bousso} R.~Bousso, Rev. Mod. Phys. 74 (2002) 825-874.

\bibitem{CGI} P.~Castorina, D.~Grumiller, A.~Iorio, \PR D77 (2008) 124034.

\bibitem{karsch} S.Datta,F.Karsch,P.Petreczky and I.Wetzorke, J. Phys. G 31 (2005) S351.

\bibitem{cea} P.Cea,L.Cosmai, F.Cuteri and A.Papa, EPJ Web Conference 175 (2018) 12006.
	
\bibitem{mannarelli1} P.Castorina and M.Mannarelli, Phys. Lett. B 644 (2007) 336.

\bibitem{mannarelli2} P.Castorina and M.Mannarelli, Phys. Rev. C 75 (2007) 054901.

\bibitem{beppe} P.Castorina, G. Nardulli and D. Zappala, Phys. Rev. D 72 (2005) 076006.

\bibitem{jap} Hu Li,C.M. Shakin and Qing Sun, Phys. Rev. D 67 (2003) 114012.

\bibitem{david} A.Wergieluk, D. Blaschke, Y.L. Kalinovsky and A. Friesen, Physics of Particles and Nuclei Letters 10 (2013) 660.

\bibitem{eddy} E. Shuryak , arXiv:1806.10487.

\bibitem{olaf} O.Kaczmarek and F.Zantow, hep/lat/0506019.

\bibitem{Lue} M.\ L\"uscher, G.\ M\"unster and P. Weisz, \NP B 180 (1981) 1.

\bibitem{terashima} H.~Terashima, \PR D61 (2000) 104016.

\bibitem{entaentropy} A.~Iorio, G.~Lambiase, G.~Vitiello, Ann. Phys. {\bf 309} (2004) 151.

\bibitem{srenidcky} M.~Srenidcky, Phys.Rev.Lett. {\bf 71} (1993) 666.

\bibitem{solodukin} S.~N.~Solodukhin, Liv. Rev. Rel. {\bf 14} (2011) 8.

\bibitem{laflamme} R.~Laflamme, Phys. Lett. B {\bf 196} (1987) 449.

\bibitem{wald} R.~M.~Wald, General Relativity, The Univ. Chicago Press (Chicago) 1984.



\bibitem{Cley} J.\ Cleymans et al., \PL B 615 (2005) 50.

\bibitem{Tawfik} A.\ Tawfik, J.\ Phys. G 31 (2005) S1105;
hep-ph/0507252 and hep-ph/050824.

\bibitem{foc} P.Braun-Munzinger and J.Stachel, Nucl. Phys.A 606 (1996) 320.

\bibitem{adriano1} A.Di Giacomo, H.G. Dosch, V.I. Shevchenko abd Yu.A. Simonov, Phys. Rep. 372 (2002) 319.

\bibitem{adriano2} M.D'Elia,A.Di Giacomo and E.Meggiolaro, Phys. Rev.D 67 (2003) 114504.


\end{thebibliography}
\end{document}